\documentstyle[12pt]{article}
\begin{document}
\parindent=0.8cm
\begin{titlepage}
\centerline{\large\bf CP violation in neutrino mixing matrix and leptogenesis}

\vspace{1cm}

\centerline{ Yong Liu$^1$ and Utpal Sarkar$^2$ }
\vspace{0.5cm}
\centerline{${}^1$ Lab of Numerical Study for Heliospheric Physics (LHP)}
\centerline{Chinese Academy of Sciences}
\centerline{P. O. Box 8701, Beijing 100080, P.R.China}
\vspace{0.5cm}
\centerline{${}^2$ Physical Research Laboratory, Ahmedabad 380 009, India }
\vspace{2cm}

\centerline{\large\bf Abstract}
\vspace{0.6cm}

The CP violation required in leptogenesis may have
different origin, but in an effective theory they all are related
to the rephasing invariant CP violating measure in the mixing
matrix of the leptonic sector. We point out that the maximum
amount of CP violation in some models can be estimated with
our present knowledge of the neutrino mixing angles, which
can help us understand the CP violation in the generation of the
lepton asymmetry of the universe. For example, the possibility
of leptogenesis may be ruled out in some models from an
knowledge of the effective neutrino mass matrix.

\vskip .3in

PACS number(s): 11.30.E, 14.60.P, 12.15.F

\end{titlepage}

Recently, the Super-Kamiokande experiment has provided a strong evidence
for non-zero masses and oscillations of neutrinos \cite{super1,super2}.
Because it is one of the direct indications for new physics beyond the
Standard Model, the announcement of the Super-Kamiokande result
has brought up a turbulent shock in the research field of particle
physics \cite{nu1,nu2}. Although
the parameter space of the neutrino mass sector or the origin of the
neutrino masses and mixing are yet to be known, these new experiments
has narrowed down the allowed parameter space in the lepton sector.

It has now become an interesting exercise to understand the allowed
parameter space in terms of different models or ansatz, as in the quark
sector, with the hope to find an origin of the neutrino masses and
mixing. There are different approches to the problem, namely, to
postulate some ansatz for simplifying the problem and then check
its consistency and predictability, or to consider some known ansatz
and check if they are consistent.
In analogy to the ansatze for the quark masses and mixing, one
can assume a similar mass and mixing matrix for the neutrino
sector. This would allow us to discuss the problem of mixing
and CP violation in neutrino system naturally
\cite{schubert,gago,barger}.

Another question of interest related to the neutrino mass is
leptogenesis \cite{lepto,triplet}. It is now believed that the most
promising mechanism for generating a baryon asymmetry of the universe
is through lepton number violation. The scale of lepton number violation
and the amount of CP violation tells us if it is possible to generate
a lepton asymmetry of the universe at the lepton number violating scale,
which then can get converted to a baryon asymmetry of the universe
in the presence of the sphalerons.
In general it is not possible to infer the amount of CP violation in
the leptonic sector, since the CP phase is an independent parameter.
However, given all the mixing angles it is possible to say what is
the maximum amount of CP violation in any model in a rephasing
invariant way. If this quantity vanishes, then one can infer that
there is no CP violation in that model and hence leptogenesis
will not be possible.

There has been several attempts to relate the various parameters
in the quark sector with an aim to understand the origin
of the quark masses and mixing. Different ansatz for the quark
masses have been put forward to reduce the number of
parameters. Some of these ans\"{a}tze has been extended to
the leptonic sector. In this article we shall study some of
these models and estimate the maximum allowed CP violation
and point out that from the study of the effective low energy
mixing matrix one can rule out the possibility of leptogenesis
in some cases.

\begin{table}
\caption{Present experimental constraints on neutrino masses and mixing}
\begin{center}
\begin{tabular}{||rcl||}
\hline \hline
Solar Neutrino {\cite{soldata1}}
&:& $\Delta m^2 \sim (0.8 - 2) \times 10^{-5} eV^2$ \\
(Large angle MSW) \phantom{[1]}
&& $ \sin^2 2 \theta \sim 1$ \\
Solar Neutrino {\cite{soldata1}}
&:& $\Delta m^2 \sim (0.5 - 1) \times 10^{-5} eV^2$ \\
(Small angle MSW) \phantom{[1]}
&& $ \sin^2 2 \theta \sim 10^{-2} - 10^{-3}$ \\
Solar Neutrino {\cite{soldata1}}
&:&$ \Delta m^2 \sim (0.5 - 6) \times 10^{-10} eV^2$ \\
( Vacuum oscillation) \phantom{[1]} &&$  \sin^2 2 \theta \sim 1$ \\
Atmospheric Neutrino \cite{super1}
&:&$ \Delta m^2 \sim (0.5 - 6) \times 10^{-3} eV^2$ \\
&&$ \sin^2 2 \theta > 0.82 $\\
Neutrinoless  Double  Beta  Decay \cite{double} &:&
$m_{\nu_e} < 0.46 eV$ \\
CHOOZ \cite{chooz} &:& $\Delta m_{e X}^2 < 10^{-3} eV^2 $\\
&&(or $\sin^2 2 \theta_{eX} < 0.2)$ \\
\hline \hline
\end{tabular}
\end{center}
\end{table}

We consider a three generation scenario with hierarchical
Majorana masses, $m_{\nu_e} \ll m_{\nu_\mu} \ll
m_{\nu_\tau}$. The neutrino masses could originate from either
see-saw mechanism \cite{seesaw} or through a triplet higgs
field \cite{triplet}. We assume
that the neutrino mass matrix is such that it can explain the
present experiments with the mass squared differences and
mixing angles as given in table 1.

We shall further assume that the solar neutrino data
is explained with $\nu_e-\nu_{\mu}$ oscillation, while the
atmospheric data can be explained in
terms of $\nu_{\mu}-\nu_{\tau}$ large mixing with a large mass splitting
compared to the $\nu_e-\nu_{\mu}$ case \cite{barbieri}.
Our result is valid when any two of the mixing angles are
given, although we need not limit which two of the three mixing
angles. For the solar neutrino problem we consider all the three possible
solutions, namely the small angle MSW, the large angle MSW and
the vacuum oscillation. The
small-mixing solution causes the energy-spectrum
distortion while the large-mixing solution causes the day-night flux
difference, and the vacuum-oscillations cause seasonal variation of
the $^7B_e$ solar neutrino flux \cite{PDG}. Since we are interested
in only the mixing angles, the large angle MSW solution and the
vacuum oscillation solution would give us same result, {\it i.e.},
they both allow same amount of CP violation.

To understand the question of CP violation in the leptonic sector,
we shall start with the neutrino mixing matrix $V_{\nu}$, which
we parametrize similar to the
standard parametrization of the Cabibbo-Kaboyashi-Maskawa
(CKM) matrix in the quark sector,
\begin{equation}
V_{KM}= \left (
\begin{array}{ccc}
   c_{12} c_{13} & s_{12} c_{13}& s_{13} e^{-i \delta_{13}} \\
   -s_{12} c_{23}-c_{12} s_{23} s_{13} e^{i \delta_{13}} &
   c_{12} c_{23}-s_{12} s_{23} s_{13} e^{i \delta_{13}}    &
   s_{23} c_{13}\\
   s_{12} s_{23}-c_{12} c_{23} s_{13} e^{i \delta_{13}}  &
   -c_{12} s_{23}-s_{12} c_{23} s_{13} e^{i \delta_{13}} &
   c_{23} c_{13}
\end{array}
\right ) .
\end{equation}
where, the convention $s_{ij}=\sin\theta_{ij}, c_{ij}=\cos\theta_{ij}$
(the "generation" labels $i,j=1,2,3$) are used. $\delta_{13}$ and
$\theta_{ij}$ are the CP phase and the mixing angles present in the
mixing matrix present in the leptonic sector. We may work in the
basis in which the charged lepton mass matrix is diagonal, in which
case this is the matrix which diagonalises the neutrino mass matrix.
With the real angles,
$\theta_{12}, \theta_{23}$ and $\theta_{13}$ can all be made to
lie in the first quadrant. The phase $\delta_{13}$ lies in the range
$0<\delta_{13}<2 \pi$. In the following, we shall fix the three angles
$\theta_{12}, \theta_{23}$ and $\theta_{13}$ in the first quadrant.

Any rephasing of the neutrino fields can change the amount of
CP violation in different sectors, but we can define a rephasing
invariant quantity, similar to the Jarlskog invariant
\cite{jarlskog,paschos} in the quark sector, given by,
\begin{equation}
\label{jarl}
J_{CP}=s_{12} s_{13} s_{23} c_{12}
c_{13}^2 c_{23} \sin {\delta_{13}}.
\end{equation}
This quantity is an measure of CP violation independent of the
basis and phases. Neutrino masses could originate from any model
which could have several sources of CP violation, but finally in
terms of this effective theory all the sources of CP violation has
to be related to this quantity $J_{CP}$.

In realistic models of  neutrino masses one can integrate out the
heavier fields (in the see-saw mechanism the right handed neutrinos
and in the triplet higgs mechanism the heavy triplet scalars ) and
get an effective low energy scenario with three generation. Then
diagonalising the charged lepton mass matrix one can obtain the
neutrino mixing matrix and hence $J_{CP}$. No matter what are the
sources of CP violation in the original model, if there is any CP
violation to start with, then the final effective model will also
violate CP and hence $J_{CP}$ has to be non-vanishing. So, if we
can predict $J_{CP}$ in any model, we can infer about the existence
of CP violation in the model. For example, to generate a lepton
asymmetry of the universe one requires CP violation. If $J_{CP}=0$
in any model, then it is not possible to generate a lepton
asymmetry of the universe in that model, no matter how complicated
the original model was.

Since the CP phase $ \delta_{13} $ is an independent parameter,
with our present knowledge it is not possible to predict $J_{CP}$.
However, with our present knowledge of the mixing matrix
$V_{\nu}$ we can compute the maximum permissible $J_{CP}$,
which we call $J_{CP}^{max}$, by choosing $ \delta_{13}=
{\pi \over 2} $. However, if we can predict $ \delta_{13} $
starting from some ansatz or some other consideration, we can
again estimate $J_{CP}^{max}$, although we may not estimate
$J_{CP}$.

Let us now consider a few specific examples. Consider the
bimaximal neutrino mixing matrix, in which $s_{13} = 0$.
In this case, $J_{CP}=0$, implying that there is no CP violation
in the leptonic sector and hence leptogenesis is impossible in
any model which produces exact bimaximal neutrino mixing
matrix. Similarly, there are models with one sterile neutrino,
where some texture neutrino mass matrix has been proposed
\cite{nu1,nu3}. In these models one has to study a $4 \times 4$
mass matrix and hence there will be three $J_{CP}$. Although
the weak mixing matrix will now be different from the neutrino
mixing matrix, one can infer that the model is CP invariant
if all the three $J_{CP}$ vanishes.
Because of the texture zeroes, all the $J_{CP}^{max}$
vanishes in a few models (which will be discussed elsewhere),
implying that although
these textures are otherwise successful, they cannot come from
any model which predicts non-vanishing lepton asymmetry of
the universe.

One may then consider a deviation from the exact bimaximal
neutrino mixing matrix and make $s_{13} \neq 0$, which will
then have CP violation. Depending on the value of $s_{13}$
the amount of CP violation will become uncertain. However,
we can then use the CHOOZ data to give an upper bound on
$s_{13}$, which will then allow us to predict $J_{CP}^{max}$
for $ \delta_{13}= {\pi \over 2} $.

>From table 1, we can use the maximum allowed value of
$\sin^2 2 \theta_{23} \sim 1$. Then for the large angle
MSW or vacuum oscillation solution of the solar neutrino
problem we can again use $\sin^2 2 \theta_{12} \sim 1$.
For the third angle we can then use the CHOOZ result,
$\sin^2 2 \theta_{13} < 0.2$ to find an experimental
upper bound on $$J_{CP}^{max}(expt) < 0.056. $$
Similarly, for the small angle MSW solution of the
solar neutrino problem we get, $$J_{CP}^{max} < 0.005. $$
For this bound we considered $\sin^2 2 \theta_{12} < .01$.
In both these cases we assumed $ \sin \delta_{13} \sim 1$.
As can be seen from these expressions for $J_{CP}^{max}$,
the amount of CP violation coming from the mixing matrix in
the leptonic sector cannot be very large.

We shall next present a model of the weak CP violation
in the quark sector \cite{cgll,liu}, which
has a geometrical origin and has got several interesting observable
predictions, which we would like to extend to the neutrino sector.
Since the amount of CP violation is predicted in this model, we
can estimate $J_{CP}^{max}$ directly. In this model,
the weak CP phase $\delta_{13}$ has been
related to the other three mixing angles $\theta_{12},
\theta_{23}$ and $\theta_{13}$ by the relation,
\begin{equation}
\label{angle}
\sin\delta_{13}=\frac{ (1+s_{12}+s_{23}+s_{13})
                       \sqrt{1-s_{12}^2-s_{23}^2-s_{13}^2+
                       2 s_{12} s_{23} s_{13}} }{(1+
                       s_{12}) (1+s_{23}) (1+s_{13})}
\end{equation}
The geometric interpretation comes from the fact that
$\delta_{13}$ is the solid angle enclosed by $(\pi/2 -\theta_{12}),
(\pi/2-\theta_{23})$ and $(\pi/2-\theta_{13})$ standing
on a same point, or, the area to which the solid angle
corresponding on a unit spherical surface.
Hence, to make $(\pi/2 -\theta_{12}), (\pi/2-\theta_{23})$ and
$(\pi/2-\theta_{13})$ be able to enclose a solid angle, the following
relation must hold.
\begin{equation}
\label{tri}
(\frac{\pi}{2}-\theta_{ij})+(
\frac{\pi}{2}-\theta_{jk}) \geq (
\frac{\pi}{2}-\theta_{ki})  \;\;\;\;\;\;
(i\not=j\not=k\not=i=1,2,3. \;\;\;\theta_{ij}=\theta_{ji})
\end{equation}
With the constraints
Eq.(\ref{tri}) and Eq.(\ref{angle}) we shall now study the predictions
of the CP violation in this scenario.

The atmospheric neutrino problem requires,
\begin{equation}
\label{pi4}
\theta_{\mu \tau}\approx \pi/4.
\end{equation}
This will give restriction on the mixing
angle between $\nu_e$ and $\nu_\tau$.  From Eq.(\ref{tri}), we have
\begin{equation}
\label{atr}
|(\frac{\pi}{2}-\theta_{e \mu})-(
\frac{\pi}{2}-\theta_{\mu \tau})| \leq (
\frac{\pi}{2}-\theta_{e \tau}) \leq
Min ( \pi/2, \;\; (\frac{\pi}{2}-\theta_{e \mu})+(
\frac{\pi}{2}-\theta_{\mu \tau}) )
\end{equation}
Note that, we can read off the mixing angles from table 1, which
implies for the small and large angle MSW solutions, to be
$$
\theta_{e \mu}\sim 0.045 \;\;\; {\rm or} \;\;\; \pi/2-0.045
$$
and
$$
\theta_{e \mu}\sim 0.7 \;\;\; {\rm or} \;\;\; \pi/2-0.7
$$
respectively.
Considering Eq.(\ref{pi4}), then we obtain
\begin{equation}
\label{etau1}
0 \leq \theta_{e \tau} \leq \pi/4+0.045
\end{equation}
or
\begin{equation}
\pi/4-0.045 \leq \theta_{e \tau} \leq  \pi/4+0.045
\end{equation}
for the case of small-mixing solution. And
\begin{equation}
0 \leq \theta_{e \tau} \leq \pi/4+0.7
\end{equation}
or
\begin{equation}
\label{etau2}
\pi/4-0.7 \leq \theta_{e \tau} \leq \pi/4+0.7
\end{equation}
for the case of large-mixing solution. Although
eq.(\ref{etau1}-\ref{etau2}) seems to be the new constraints in this
scenario, they are irrelevant. Considering the CHOOZ data we can
easily see that the region allowed in this scenario is just the region
allowed by CHOOZ,
\begin{equation}
0 \leq \sin \theta_{e \tau} \leq .23.
\end{equation}

Substituting eqn (\ref{angle}) into eqn (2), we obtain
$J_{CP}$ as a function of
$\theta_{e \tau}$ and can draw the curve with $J_{CP}$ versus
$\theta_{e \tau}$. The results are shown in fig. 1.
>From the figure we can put a limit on the
amount of CP violation, from the limit on $\theta_{e \tau}$ to be,
\begin{equation}
J_{CP} < 0.0015.
\end{equation}
However, there is another direct way to give bound on the amount of
CP violation in this scenario. For this we assume that $J_{CP}$
corresponds to the largest value of $\delta_{13}$ (which we can
verify from the graph).

Using eq.(\ref{angle}) we can get an
upper bound on the CP phase in this parametrization to be
$\sin \delta_{13} \sim .13$ for the large angle solution of the
solar neutrino problem, so that $\sin^2 2 \theta_{e \mu} \sim 1$ and
$\sin^2 2 \theta_{\mu \tau} \sim 1$ and maximum allowed value for the third
angle to be given by CHOOZ, $\sin^2 2 \theta_{e \tau} < .2$. These values
will then give a maximum allowed value for the rephasing invariant
CP violating qantity,
\begin{equation}
J_{CP}^{max} < 0.0175 .
\end{equation}
On the other hand for the small angle MSW solution the CP phase
is predicted to be larger, $\delta_{13} \sim 0.61$ and the
rephasing invariant CP violating quantity becomes
becomes
\begin{equation}
J_{CP}^{max} < 0.0034 .
\end{equation}
Again we obtain a maximum measure of CP violation using the
largest value of $\sin^2 2 \theta_{12} \sim 0.01$.
In both the cases the amount of maximum CP violation
is much lower than the experimental bounds.

To summarise, we have shown that it is possible to estimate
the maximum allowed value of the rephasing invariant measure
of CP violation in a given model if we know all the three angles. Since
the CP phase is an independent parameter, one can assume a  maximum
value of unity for this quantity to calculate the rephasing invariant
CP violating parameter. In exactly bimaximal mixing model and in
some models of textured neutrino mass matrix, this
measure $J_{CP}^{max}$ vanishes implying no CP violation,
whereas in another geometric model of weak CP phase there is
a large suppression. Any ansatz of the neutrino mixing matrix
can, in general, suppress this quantity, which in turn will suppress
the amount of CP violation in that model, whose direct effect
will be on the amount of lepton asymmetry of the universe.
In models with
$J_{CP}^{max}=0$, it is not possible to have lepton asymmetry
of the universe.

\newpage
\begin{figure}[htb]
\mbox{}
\vskip 6.5in\relax\noindent\hskip -.6in\relax
\includegraphics{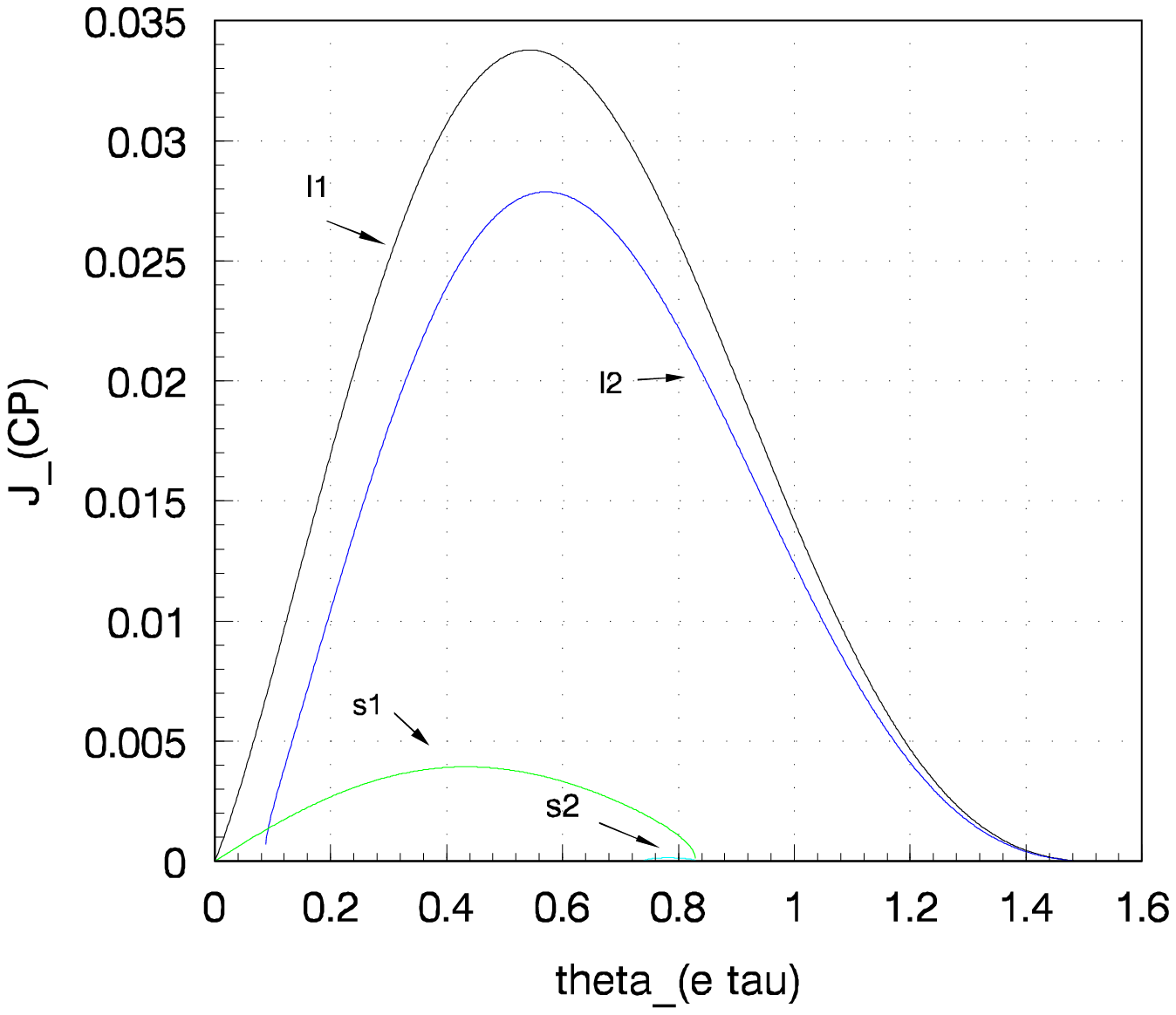}
\caption{ $J_{CP}$ versus $\theta_{e \tau}$. Where,
$\theta_{\mu \tau}=\pi/4$. The curves s1, s2, l1 and l2 corresponds to
the cases of
$\theta_{e \mu}=0.045,\; (\pi/2-0.045),
\; 0.443$ and $(\pi/2-0.443)$ respectively.}
\end{figure}

\end{document}